\documentclass[12pt]{article}
\usepackage{graphicx}
\usepackage{amssymb}
\usepackage{cite}
\usepackage{bm}
\begin{document}
\setlength{\unitlength}{1mm}
\textwidth 15.0 true cm 
\headheight 0 cm
\headsep 0 cm 
\topmargin 0.4 true in
\oddsidemargin 0.25 true in
\input epsf

\newcommand{\beq}{\begin{equation}}
\newcommand{\eeq}{\end{equation}}
\newcommand{\be}{\begin{eqnarray}}
\newcommand{\ee}{\end{eqnarray}}
\renewcommand{\vec}[1]{{\bf #1}}
\newcommand{\vecg}[1]{\mbox{\boldmath $#1$}}
\newcommand{\grpicture}[1]
{
    \begin{center}
        \epsfxsize=200pt
        \epsfysize=0pt
        \vspace{-5mm}
        \parbox{\epsfxsize}{\epsffile{#1.eps}}
        \vspace{5mm}
    \end{center}
}

\begin{flushright}

\end{flushright}

\vspace{0.5cm}

\begin{center}

{\Large\bf   Benign vs. malicious ghosts in higher-derivative theories}

\vspace{1cm}

{\Large A.V. Smilga} \\

\vspace{0.5cm}

{\it SUBATECH, Universit\'e de
Nantes,  4 rue Alfred Kastler, BP 20722, Nantes  44307, France. }
\footnote{On leave of absence from ITEP, Moscow, Russia.}\\

\end{center}

\bigskip

\begin{abstract}
 Interacting theories with higher-derivatives involve ghosts. 
They correspond to instabilities that
display themselves at the classical level. 
We notice that comparatively ``benign'' mechanical higher-derivative systems exist,
where the classical vacuum is stable with respect to small perturbations and the problems appear only
at the nonperturbative level. We argue the existence of benign 
higher-derivative field theories  which are stable with respect to small fluctuations with nonzero
momenta. A particular example is the $6D$ ${\cal N} =2$  higher-derivative SYM theory, which is finite and unitary
at the perturbative level.
The  instability with respect to small fluctuations of zero-momentum modes 
is always present, however.    
\end{abstract}

\section{Motivation.}
There are two common contexts where the theories with higher-derivative terms in the lagrangian
are usually considered: {\it (i)} they appear in effective low--energy lagrangians;
{\it (ii)} they can be introduced to regularize theory in the ultraviolet.
If treating such a theory as a fundamental one, ghosts appear. Ghosts usually spoil unitary
and/or causality of the theory and this is the reason
by which higher-derivative theories are {\it not} usually considered as candidates for the 
Theory of Everything.
  
We  think, however, that, ghosts nontwithstanding, the idea to have a higher-derivative
theory at the fundamental level is not altogether stupid, and one should continue to think 
in this direction. There are two considerations which make this idea
rather suggestive. 

\begin{itemize}
\item The first one comes from the half-a-century efforts to quantize gravity.
 It is well known that the ordinary Einstein gravity involves a dimensionful coupling 
constant and is not renormalizable.
This refers also to supergravity, even the maximally extended one. 
On the other hand, the (quartic in the derivatives of the metric)  action
  \be
\label{Weylgrav}
  S \ =\  \frac 1{2\alpha}  \int  
C_{\mu\nu\rho\sigma} C^{\mu\nu\rho\sigma}
\sqrt{-g} \ d^4 x \ ,
  \ee
where
 \be
\label{Weyltens}
C_{\mu\nu\rho\sigma} \ =\ R_{\mu\nu\rho\sigma} +
\frac 12 \left[ g_{\mu\sigma} R_{\nu\rho} + 
 g_{\nu\rho} R_{\mu\sigma} -  g_{\mu\rho} R_{\nu\sigma}
-  g_{\nu\sigma} R_{\mu\rho} \right] \nonumber \\
+ \ \frac R6 \left[  g_{\mu\rho}  g_{\nu\sigma} -
 g_{\mu\sigma} g_{\nu\rho} \right]
   \ee
is the Weyl tensor, 
involves a dimensionless coupling constant $\alpha$ and is renormalizable. The same
is true in conformal supergravity. The ${\cal N} =1$ conformal supergravity enjoys
asymptotic freedom, while a version of  ${\cal N} =4$ conformal supergravity is
{\it finite} so that the symmetry of the classical  action
(\ref{Weylgrav}) with respect to local conformal transformations
$$ g_{\mu\nu}(x) \ \to \ \lambda(x) g_{\mu\nu}(x) $$
is also the symmetry of the quantum theory \cite{Fradkin}. 
Conformal gravity has also a comparatively
benign behavior in the non--perturbative region. At least, the 
Euclidean action is positive
definite there (in contrast to the Einstein case). This allows one to give a sensible
physical intepretation to gravitational instantons. Similar to usual BPST Yang--Mills instantons,
the Eguchi--Hanson instanton can be interpreted as a tunneling trajectory connecting flat vacua with
opposite orientation \cite{jainst}. A more detailed discussion of this and related questions
was given in recent \cite{dragon}. One can add that conformal supergravity 
 pops up as an 
effective action for the twistor--string theory, which attracted recently a considerable 
interest \cite{Witconf}.

\item Another  reason by which higher-derivative theories may be attractive  is also
associated with  attempts to quantize gravity. In four dimensions this does not work not
only because of nonrenormalizability, but also due to a  known conceptional difficulty: 
any theory, like gravity, where no universal {\it flat} time can be introduced  is 
intrinsically acausal, which makes its quantization problematic \cite{noncause,dragon}.
A rather popular modern idea is that the world has actually more than four dimensions 
and our Universe represents a thin curved $4D$ brane embedded in  flat higher-dimensional 
bulk \cite{RS}. Then one should understand what is the fundamental theory in the bulk. If it is a 
{\it field theory}\, \footnote{A modern paradigm is that the TOE is 
not a field theory, but rather
a version of string theory. However, in spite of intense studies during the last 20 years, we
still lack a consistent non--perturbative formulation of the latter.}, 
it cannot be an ordinary theory, with the 
 lagrangian containing at most two derivatives. All such theories involve a dimensionful coupling
in higher dimensions 
and are not renormalizable. Note that nonrenormalizability is a more
serious trouble than just difficulties in calculating the perturbative series. 
Nonrenormalizability also 
means that the continuous limit of the path integral does not exist and quantum theory simply
cannot be {\it defined}. On the other hand, a higher-derivative higher-dimensional 
theory may involve a dimensionless coupling and be renormalizable. The simplest 
example is a $6D$ YM theory with the action
\footnote{Let us comment about the sign in Eqs. (\ref{Weylgrav}) (\ref{FBoxF}). 
In Minkowski space with path integral
$\sim \int \exp\{iS_M\}$, the choice of sign for $S_M$ is a pure convention. 
We could try to  fix it by requiring that, after the usual
Euclidean rotation $t \to -i\tau$ is done, the path integral 
has the form $\int \exp\{-S_E \}$, where the Euclidean
action $S_E$ is positive definite for small fluctuations. This would give the {\it negative}
 signs   in Eqs. (\ref{Weylgrav}) (\ref{FBoxF}). Unfortunately, we will see that, 
in supersymmetric higher-derivative models we are about to consider, the Euclidean action cannot be made
positive definite: the contributions of the fields of different spin have different signs.
Thus, we have done another choice of sign based on the requirement for 
 the Minkowskian highest-time-derivative kinetic term to be positive.  In contrast to what happens in conventional
theories, it {\it does} not change sign after Euclidean rotation.}   
  \be
 \label{FBoxF}
 \frac 1{2g^2} {\rm Tr} \int d^6x \, F_{\mu\nu} D^2 F_{\mu\nu}.       
   \ee

\end{itemize}

As was mentioned, the theories like (\ref{Weylgrav}, \ref{FBoxF}) involve
ghosts.
This paper represents a simple observation concerning the nature 
of these ghosts. Note first of
all that ghosts are not specific for quantum field theory. They are 
seen also in quantum mechanics.
Moreover, the origin of the ghosts is clearly seen at the classical 
level. 
It turns out that interactive
dynamic systems with higher derivatives in the lagrangian involve 
instabilities. When the system
is quantized, these instabilities make Hamiltonian non--Hermitean 
and $S$--matrix non--unitary (see e.g. a recent discussion in \cite{Hawlive}).  
The instabilities can be, however, of two types: the perturbative 
{\it malicious} instabilities
and comparatively {\it benign} nonperturbative ones. A perturbative 
instability is the instability 
of vacuum with respect to small fluctuations. Whenever it is present, 
perturbation theory
 makes little sense, which is troublesome. On the other hand, in a 
theory involving only 
benign instabilities, the vacuum state is metastable and  problems appear 
only when the amplitude 
of the fluctuations reaches some threshold. In the vicinity of vacuum, 
the classical trajectories are smooth
and regular. On the other hand, there is a region in the phase space, where the 
trajectories become singular. 
This phenomenon is well known in usual quantum (and classical !) 
mechanics with a strong attractive 
  potential and is called { falling on the center} or {\it collapse}.
Note that processes with black hole formation 
 in conventional Einstein gravity have exactly the same nature.

Most higher-derivative systems have malicious ghosts. However, in the 
next section we present a classical mechanical model which is quartic in
derivatives and involve only benign nonperturbative ghosts.   Field theories
are discusses in Sect. 3. We show that a theory like (\ref{FBoxF}) is {\it almost} benign:
the vacuum is stable there with respect to  fluctuations with a finite wavelength and is unstable
only with respect to certain {\it homogeneous} field fluctuations. 
One can speculate on the relevance of this observation for the inflation scenario.  
We dwell on a special class of theories, the  finite conformal
$6D$ theories. They have many attractive features and make the strongest bid to 
enjoy a  vacuum state that is stable with respect to small fluctuations of the modes with nonzero momenta. 
(Instability with respect to the fluctuations of the zero-momentum mode is
a family feature of all such theories.)
 Finally, we  briefly discuss conformal supergravity and notice that ${\cal N} =4$ 
conformal supergravity  is stable in the  linearized approximation.

\section{Toy models.}
Consider the lagrangian
 \be
\label{L2}
{\cal L} \ =\  \frac 12 \ddot q^2 -   \frac {\Omega^4}2 q^2 \ .
 \ee
The canonical equations of motion for a  lagrangian involving second derivatives is
 \be
\label{caneqmot}
\frac {d^2}{dt^2} \frac {\delta {\cal L}}{\delta \ddot q} - 
\frac {d}{dt} \frac {\delta {\cal L}}{\delta \dot q} + 
\frac {\delta {\cal L}}{\delta  q} \ =\ 0 \ .
 \ee
For (\ref{L2}) this gives $q^{(4)} - \Omega^4 q = 0$. 
The characteristic equation $\lambda^4 - \Omega^4 = 0$
has both real and imaginary solutions. The latter 
give conventional oscillatory solutions
and the former --- exponential solutions 
$\sim \exp\{\lambda t\} = \exp\{\pm \Omega t\}$. 
The exponentially rising mode
signalizes the instability of the vacuum
\footnote{We have to comment on what {\it vacuum} in higher-derivative theories means. In usual theories,
classical vacuum is a static solution to the equations of motion with minimal energy. However, we will shortly
see that, in  higher-derivative theories, the energy functional is not bounded from below (neither it is bounded 
from above) and 
vacuum will be simply understood as a static classical solution. 
In field theories, we will call vacuum a spatially homogeneous
static solution.}
 $q=0$ and {\it is}  
a malicious ghost in our terminology.

 Consider now another system,
\be
\label{L4}
{\cal L} \ =\ \frac 12 (\ddot q  + \Omega^2 q )^2 -  \frac {\alpha}4 q^4 \ .
 \ee
The corresponding equations of motion are
 \be
\label{eqmot4}
\left( \frac {d^2}{dt^2} + \Omega^2 \right)^2 q - \alpha q^3 \ =\ 0\ .
 \ee
Note first of all that in the linear approximation, $\alpha = 0$, the characteristic
equation $(\lambda^2 + \Omega^2)^2$ has degenerate imaginary roots. This may give linearly rising
with time solutions, but not exponentially rising ones. The nonlinear term $\propto \alpha$ brings
about instabilities, however. To understand why it does, recall first why in many conventional
dynamic nonlinear systems, there is no instability. The point is that the energy integral
$\dot q^2/2 + V(q)$ represents a sum of the positive definite kinetic term and the potential.
If the latter is bounded from below, the kinetic energy is  bounded   
from above and so is $|\dot q|$.

The equation of motion (\ref{caneqmot}) has also a conserved energy integral \cite{Ostr}
 \be
E \ =\ \left( \ddot q  -  \dot q  \frac d{dt} \right) 
  \frac {\delta {\cal L}}{\delta \ddot q} + \dot q
 \frac {\delta {\cal L}}{\delta \dot q} -  {\cal L}\ ,
 \ee
but this does not restrict the derivatives $\dot q, \ddot q, q^{(3)} $ to be arbitrary high.
For the lagrangian (\ref{L4}), the energy is 
 \be
E \ =\ \ddot q  (\ddot q + \Omega^2 q ) - \dot q  (q^{(3)} + \Omega^2 \dot q )
- \frac 12  (\ddot q + \Omega^2 q )^2 + \frac {\alpha}4 q^4 \ .
 \ee 
There are the terms of different sign and a  reachable by classical trajectories 
corner in the phase space exists, where the derivatives grow exponentially
and even much faster than that, reaching singularity in a finite time. 
This is the collapse phenomenon. Actually, one can just {\it observe}
this singularity solving the equation (\ref{eqmot4}) with {\sl Mathematica}. One can formulate 
a conjecture:

{\sl Any nonlinear lagrangian system involving higher derivatives can collapse}.
In other words, some of the classical trajectories can hit a singularity
\footnote{We do not know whether this conjecture can be (or maybe already has been) proven
by mathematicians.}

Our second crucial observation is that though the system {\it can} collapse, it does not
{\it have to}. The equation (\ref{eqmot4}) admits benign regular orbits in the vicinity
of the stationary point $q = \dot q = \ddot q = q^{(3)} = 0$. Only fluctuations of large
enough amplitude go astray. There is a separatrice between the benign perturbative
and the wild nonperturbative regions. If posing the initial conditions
 $$
q(0) = c;\ \ \dot q(0) =  \ddot q(0) = q^{(3)}(0) =  0 
 $$
and turning the computer on, one can find that the threshold amplitude is 
$c_{\rm crit} \approx  0.3 \, \Omega^2/\sqrt{\alpha}$. The dependence on $\Omega$ and $\alpha$ 
follows, of course, from simple scaling arguments.  

\begin{figure}[h]
   \begin{center}
 \includegraphics[width=4.0in]{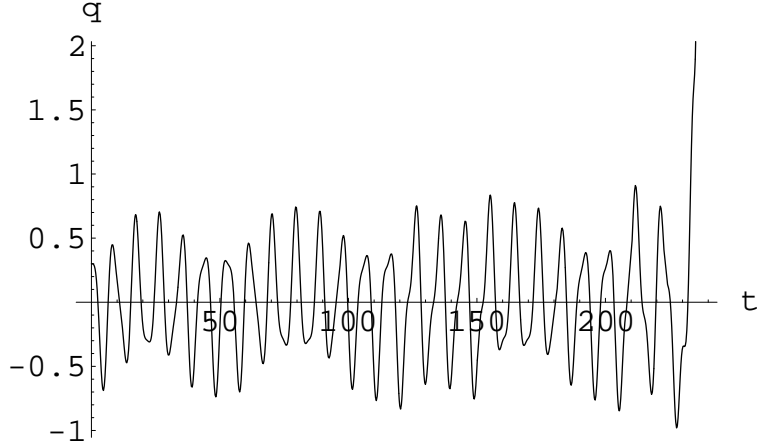}
        \vspace{-2mm}
    \end{center}
\caption{\small Oscillating and collapsing}
\label{traduh}
\end{figure}

For illustration, we plotted in Fig.\ref{traduh} the solution to the equation (\ref{eqmot4})
for $\Omega = \alpha = 1$ and $q(0)$ just above $c_{\rm crit}$. After some quasiharmonic oscillations,
the solution finally goes astray and hits the singularity. For $q(0) < c_{\rm crit}$, it keeps oscillating.
 
Note that the positive sign of $\alpha$ is crucial for such a restricted stability to take place. If the kinetic
and the potential term in the lagrangian have the same sign, the vacuum is always 
unstable with respect to small fluctuations,
though it is not a Lyapunov type of instability, seen for linearized equations.

\begin{figure}[h]
   \begin{center}
 \includegraphics[width=3.0in]{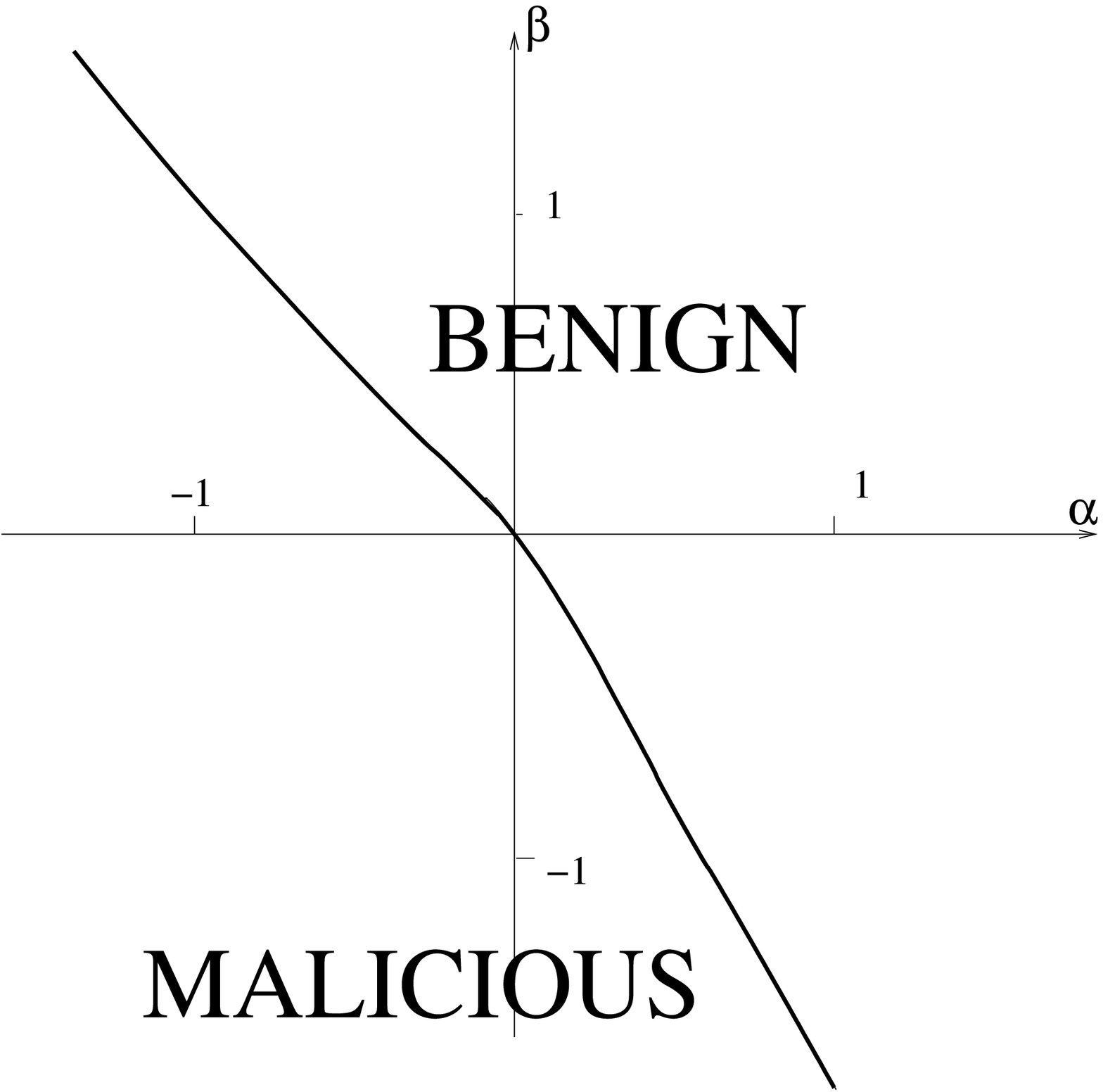}
        \vspace{-2mm}
    \end{center}
\caption{\small Benign and malicious parameter regions}
\label{phase}
\end{figure}

Another  example is
  \be
\label{L4dot}
{\cal L} \ =\ \frac 12 (\ddot q  + \Omega^2 q )^2 - \frac {\beta}2 q^2 \dot q^2 \ .
 \ee
Also for this system
 the trajectories are stable, provided $\beta$ is positive and 
$q(0) \leq 0.47 \Omega/\sqrt{\beta}$. They collapse for larger values of $q(0)$. 


Our final example (important for the future discussion) is   
\be
\label{L4mix}
{\cal L} \ =\ \frac 12 (\ddot q  + \Omega^2 q )^2 -  \frac \alpha 4 q^4 - \frac {\beta}2 q^2 \dot q^2 \ .
 \ee
The lagrangian involves  two kind of nonlinear terms: $\sim q^4$ and $\sim q^2 \dot q^2$. 
We have seen that  the system is benign if the nonlinear terms  in the lagrangian have opposite sign, compared to the
quadratic one. Thus, we expect that the system is benign if both $\alpha$ and $\beta$ are positive and malicious
if  both $\alpha$ and $\beta$ are negative. If $\alpha$ and $\beta$ have opposite signs, it depends. A phase diagram
in the $(\alpha, \beta)$ plane that reveals benign and malicious regions of parameters is drawn in Fig.\ref{phase}.

\section{Field theories.} 

 In the effective theories usually considered, 
 higher-derivative terms in the lagrangian appear as an admixture to the 
standard kinetic term. If treating
such a theory as a fundamental one, ghosts of malicious variety appear. Indeed, consider
the lagrangian 
 \be
\label{primes}
{\cal L} \ =\ - \frac 12 \phi \Box \phi  + \frac \kappa 2  \phi \Box^2 \phi - \frac {m^2}2 \phi^2
+\ {\rm interactions}\ .
 \ee
The dispersive equation $k^2 + \kappa k^4 = m^2$ has complex solutions, signalizing
 vacuum instability. This is true also for the lagrangian $\phi \Box^2 \phi/2  - m^2 \phi^2/2$
involving only the higher-derivative term and the mass term. Consider, however, the massless 
lagrangian
   \be
\label{L4fi}
{\cal L} \ =\  \frac 12   \phi \Box^2 \phi - \frac {\alpha}4 \phi^4\ .
 \ee
If suppressing the interaction term, the dispersive equation $(\omega^2 - \vec{k}^2)^2 = 0$ has
only real solutions as was the case for the model lagrangian (\ref{L4}) without the quartic term.
In fact, the interactive lagrangian (\ref{L4}) can be thought of as the field theory lagrangian
(\ref{L4fi}) restricted on a sector with given momentum $\vec{k}$. 
\footnote{Of course, this is not exact. The lagrangian (\ref{L4fi}) involves the terms 
describing interactions between the modes with different momenta. But a qualitative 
behavior of the two lagrangians is the same.} 
We see that the momentum $|\vec{k}|$ plays the role of the dimensionful parameter $\Omega$
in (\ref{L4}). We have learned that the modes with the amplitude below a threshold, 
proportional to $ \Omega^2 =  \vec{k}^2$, are stable, but there are unstable
 modes with the amplitude exceeding this threshold. 
On the other hand, vacuum {\it is} unstable
with respect to small   fluctuations of the constant mode. This is an ``almost benign'' situation.

  There might be  certain cosmological implications of this observation. The inflation scenario
 implies the {\it coherent} initial conditions where  the {constant} mode of a certain scalar field has 
a large nonzero value, while the modes with nonzero
momentum are suppressed. 
It is not trivial, however, to suggest a natural mechanism providing for such coherence. 
We see that, in the  theory (\ref{L4fi}), the field becomes, indeed, more and more coherent
as the instability develops. Of course, very soon this quasicoherent field becomes singular, and
 the situation is no longer under control.

The lagrangian (\ref{L4fi}) is massless, but this masslessness is accidental and disappears 
when radiative corrections are taken into account. 
Let us ask two questions: 
  \begin{enumerate}
\item  Do there exist naturally massless theories with ``almost benign'' ghosts ?  
\item  Do there  exist   theories with {\it exclusively} benign ghosts such that  vacuum
is stable with respect to small fluctuations with {\it all} momenta including $\vec{k} =0$ ?
  \end{enumerate}
The answer to the first question is positive and to the second --- probably negative.

 A simple observation is 
that there is a class of theories where certain scalar v.e.v.'s {\it play the role} of momenta and
may stabilize the ghosts.
 Recall what happens in ${\cal N} = 4$ SYM theory. It involves six real adjoint scalar
fields $\phi_i$. The potential $\sim \sum_{ij} {\rm Tr} \, [\phi_i, \phi_j]^2 $ vanishes
if $ [\phi_i, \phi_j] =0$ for all $i,j$. This gives an infinite set of classical vacua
(the vacuum valley or vacuum moduli space ). Using  gauge freedom, all $\phi_i$ can be put 
into the Cartan subalgebra. For $SU(2)$ this gives $\phi_i = c_i\tau^3$ and the valley
is parametrized by six real numbers $c_i$.  
Supersymmetry dictates that the degeneracy along the valley
 is not lifted after quantum corrections are taken
into account.

 The  ${\cal N} = 4$ SYM theory has the vanishing $\beta$ function. Thus, classical
conformal symmetry is not broken by quantum effects and the theory does not have an
intrinsic scale parameter. However,  scale parameters are introduced by 
the dimensionful v.e.v.'s $c_i$ .  
Recall now that the ${\cal N} = 4$ $4D$ SYM theory can be obtained by dimensional reduction from
the ten--dimensional one. In ten dimensions, $\phi_i$ played the role of extra dimensional
components of the $10D$ gauge potential $A_M$. And 
we know that gauge potentials
``come together'' with momenta in the lagrangian. 

 For sure, ${\cal N} = 4$ $4D$ SYM is a standard quadratic theory, 
not the one we need. Our point is 
that a six--dimensional quartic in derivatives SYM theory exists whose properties are similar
to those of  quadratic   ${\cal N} = 4$ one. In particular, 
 \begin{itemize} 
\item  It is finite.
 \item  It involves
the vacuum valley associated with the v.e.v.'s of scalar fields.
\item  The theory 
can be obtained by the dimensional reduction from ten dimensions such that 
the scalar fields parametrizing the vacuum
appear as extra dimensional components of the $10D$ gauge potential  and 
``come together'' with momenta in the lagrangian.
 \end{itemize}

We are talking here about  ${\cal N} = 2$ (in six-dimensional sense; it gives a  ${\cal N} = 4$ theory
in four dimensions) $6D$ higher-derivative SYM theory
 \footnote{ To the best of our knowledge, 
this theory has never been studied. 
 Even conventional $6D$ supersymmetric gauge theories did not attract an enormous attention of theorists
 and almost nothing can be found in the literature about higher-derivative $6D$ theories.}. 
We will describe now how its lagrangian looks like, display the presence of the vacuum valley   and argue 
that   $\beta$ function  vanishes there.

Supersymmetric lagrangians are best written in superfield language. Unfortunately,  ${\cal N} =1$  
$6D$ superfields are more
complicated than ${\cal N} =1$ superfields in four dimensions. They are related to   
 ${\cal N} =2$  
$4D$ superfields and are best
 described in harmonic superspace formalism \cite{hss}. For $D=6$,
this was implemented in \cite{Zupnik}. Let us, however, think in 4--dimensional terms and write
some higher-derivative  ${\cal N} =2$  supersymmetric lagrangians in four dimensions. First, let the theory
be Abelian. The lagrangian of the pure  ${\cal N} =2$  $4D$ photodynamics with higher derivatives is easily
written as
  \be
\label{phothder}
 {\cal L} =  \frac 1{2e^2} \int d^8\theta \, \bar {\cal W}  {\cal W}\ ,
 \ee
where the  ${\cal N} =2$ chiral superfield ${\cal W}$ is expressed via the 
conventional  ${\cal N} =1$ superfields
$\Phi (x_L, \theta) $ and $W_\alpha (x_L, \theta) $ as
  \be
 \label{Wcal}
{\cal W} (x_L, \theta_\alpha, \tilde{\theta}_\alpha) 
\ =\ \Phi + i \sqrt{2} \tilde{\theta}^{\alpha} W_\alpha - \frac {\tilde{\theta}^2}4 
\bar D^2 \, \bar \Phi \ . 
  \ee 
The component form of the lagrangian is
 \be
\label{phhdercomp}
e^2  {\cal L} = \frac 14 F_{\mu\nu} \Box  F_{\mu\nu} + \bar \phi \Box^2 \phi 
- \frac 12 D \Box D - \bar F \Box F - i(\sigma_\mu)_{\alpha \dot\beta}
  \sum_{f=1,2} \, \lambda^\alpha_f \partial_\mu \Box \bar \lambda^{\dot\beta}_f\ .
 \ee
Note that the former auxiliary fields $D,F,\bar F $ become dynamical. The leading-order kinetic terms
for the spin 1 field $F_{\mu\nu}$ and for the spin 0 fields $D,F,\bar F$ have the same sign in the Minkowski
lagrangian. Note, however, that after Euclidean rotation the signs become {\it different}.
 
  The coupling of the gauge sector with the matter hypermultiplet can be written
in terms of the analytic superfield $q^+(x,\theta^+, \bar\theta^+,  u)$ , where $\theta^+_\alpha = u_i^+ \theta^i_\alpha, \ \bar\theta^+_\alpha = u_i^+  \bar\theta^i_\alpha$
and $u_i^+,\ i=1,2$, are harmonic variables  parametrizing  $CP^1 \equiv S^2$. (See \cite{hss}
for details). For the conventional ${\cal N} =2$ QED, the coupling is written ( cf. Eq.(7.17) of
Ref.\cite{hss} ) as
\be
\label{717}
{\cal L}_{\rm matter}^{{\cal N} =2} \ =\ - \int du \,  d^4\theta^+ \,    
\stackrel{\smile}{{q}^+} \nabla^{++}  q^+  \ ,
  \ee
where  $ \nabla^{++} = 
D^{++} + iV^{++}$,  
 $D^{++}$ is a harmonic derivative (in a certain, so called central basis it is just 
$D^{++} = u^{+i} \frac \partial {\partial u^{-i}}$,\ $u^{-i}$ being the complex conjugate of   
$-u^+_i$), the ``smile'' operator $\smile$ is a special analiticity preserving conjugation
(if $q^+ = \phi^i u_i^+ + \ldots$, then $\stackrel{\smile}{q^+} = \bar \phi_i u^{i+} + \ldots$ )   
and $V^{++}$ is an analytic superfield related
to ${\cal N} =2$ chiral superfield ${\cal W}$ roughly in the same way as the ${\cal N} =1$ superfield
$V$ relates to $W_\alpha$. A possible choice for the higher-derivative matter term is
   \be
\label{QEDhder}
{\cal L}_{\rm matter}^{{\cal N} =2}({\rm high\ der}) \ =  \ -\frac 12  \int du  \, d^8 \theta \,   
\stackrel{\smile} {q^+} (\nabla^{--})^2 \nabla^{++}  q^+  \ ,
  \ee
where  $ \nabla^{--} = D^{--} + iV^{--}$ and 
$$
V^{--}(X, u) \ =\ \int dv \frac {V^{++}(X,v)}{(u^+ v^+)^2}\ .
$$
Note that the integral in Eq.(\ref{QEDhder}) is written over the whole superspace, 
not over its analytic subspace
and this gives extra dimension $\sim m^2$, provided by the box  operator in the component
language. Note also that  a more simple trial expression for the action 
  \be
\label{718}
{\cal L}_{\rm matter}^{{\cal N} =2} \ \sim \  \int du \,  d^8\theta \,    
\stackrel{\smile}{{q}^+} \nabla^{--}  q^+  \ ,
  \ee 
would not work as such. Indeed, the kinetic part of (\ref{718}) just vanishes (cf. Eq.(8.78) of
Ref.\cite{hss}):
$$
\int d^8\theta \stackrel{\smile}{{q}^+} D^{--}  q^+ = 
\int d^4\theta^+ (D^+)^2 (\bar D^+)^2 \stackrel{\smile}{{q}^+} D^{--}  q^+ = \ 0
$$
due to  analiticity of $q^+$ and $\stackrel{\smile}{q^+}$, $D^+_\alpha q^+ = \bar D^+_{\dot{\alpha}}
q^+ = D^+_\alpha \stackrel{\smile} {q^+} = \bar D^+_{\dot{\alpha}}
\stackrel{\smile}{q^+} =  0$,  and to commutation relations
$$ [D^+_\alpha, D^{--}] = D^-_\alpha,\ \ \ \ \ \ \{D^+_\alpha, D^-_\beta \} = 0\ .$$  

The interaction term in Eq.(\ref{718}) does not vanish, however. We will show in Ref.\cite{ISZ} that
the  requirement for the action to be conformally invariant  distinguishes a certain linear combination of
Eq.(\ref{QEDhder}),   Eq.(\ref{718}) and of the terms involving higher powers of harmonic derivatives: 
  \be
 \label{summa}
{cal L} \ =\   \int du  \, d^8 \theta \, \sum_{n=0} c_n   
\stackrel{\smile} {q^+} (\nabla^{--})^{n+1} ( \nabla^{++})^n  q^+ 
  \ee

Adding (\ref{phothder}) and (\ref{summa}) and lifting this up to 6 
dimensions (one can do it at the component
level or else use the formalism of Ref.\cite{Zupnik} to obtain the explicit $6D$ 
harmonic superspace expressions \cite{ISZ} ) , one obtains
an Abelian $6D$ gauge theory with  dimensionless coupling. Its perturbative properties are  similar
to those of the conventional QED. It is renormalizable. The one--loop renormalization of the
effective charge is
\footnote{The calculation of the numerical coefficient $c$ is under way now. A positive sign of $c$ corresponds to the Landau
pole situation and negative to asymptotic freedom. Very preliminary results \cite{ISZ} suggest that the second possibility is realized.}  
  \be
\label{e6Drenorm}
 \frac 1{e^2(\mu)} \ =\ \frac 1{e^2_0} + \frac c{64\pi^3} \ln \frac \Lambda \mu\ .
 \ee  
It is not so trivial to write down in superfields a non-Abelian ${\cal N} =2$ $4D$ higher-derivative 
theory. The problem is that, in non--Abelian case, the superfield ${\cal W}$ is not gauge invariant, 
but transforms as 
$${\cal W} \to e^{i\Lambda} {\cal W}  e^{-i\Lambda} \ \ \ \ {\rm and} \ \ \ \ 
  \bar {\cal W} \to e^{i\bar \Lambda} \bar {\cal W}  e^{-i\bar\Lambda}\ .  $$
A naive generalization of (\ref{phothder}) with $ \bar {\cal W}  {\cal W} \to \ {\rm Tr} 
\left\{  \bar {\cal W}  {\cal W} \right \}$ 
is not gauge invariant. To make it gauge invariant, one should introduce
the ``bridge'' superfield $b$ which is transformed as $e^{ib} \to e^{i\Lambda} e^{ib}  e^{-i\bar\Lambda}$
and write
   \be
\label{SYMhder}
 {\cal L} =  \frac 1{g^2} {\rm Tr} \int d^8\theta \, \left \{ \bar {\cal W} e^{-ib}  {\cal W} e^{ib} \right \}\ .
 \ee
In ${\cal N} =1$ case, this bridge is none other than the vector superfield $V$. 
In  ${\cal N} =2$ case, we meet a  problem  that no simple explicit expression 
of the bridge via the harmonic potential $V^{++}$ is known.
\footnote{The problem is there in $4D$, but, remarkably, an explicit superfield expression for the higher-derivative 
action can be written in six dimensions \cite{ISZ}.}

Anyway, the theory exists and can be explicitly 
written in terms of ${\cal N}=1$ superfields and in components. 
The part of the action depending only on ${\cal N} =1$ gauge superfields has the form
 \be
 \label{N1SYMhder}
 \frac i{g^2} (\sigma_\mu)_{\alpha \dot\beta} {\rm Tr} \, \int d^4\theta \ 
\left\{ e^V W^\alpha e^{-V} \nabla_\mu \bar W^{\dot\beta} \right\}\ ,
   \ee
where $W_\alpha = (1/8) \bar D^2 \left( e^{-V} D_\alpha e^V \right) $ and
$$\nabla_\mu \ =\ - \frac i4 (\bar \sigma_\mu)^{\dot\beta \alpha} 
\left[ e^{-V} D_\alpha e^V \bar D_{\dot\beta} + \bar D_{\dot\beta} \left(e^{-V} D_\alpha e^V \right)
\right] $$
is the covariant derivative. Lifting this up to $6D$, we obtain a  renormalizable 
theory with  dimensionless coupling constant. 
In contrast to the Abelian $6D$ theory considered above and very much similar to what is known
for $4D$ non-Abelian gauge theories, the non-Abelian $6D$ ${\cal N} =1$ theory is asymptotically free.
 We have not yet calculated the $\beta$ function  accurately, but
rather suggestive arguments lead to the following result for the effective
coupling
    \be
\label{g6Drenorm}
 \frac 1{g^2(\mu)} \ =\ \frac 1{g^2_0} - \frac c{32\pi^3} \ln \frac \Lambda \mu
 \ee  
 with the same coefficient $c$ as in Eq.(\ref{e6Drenorm}) 
(we restrict ourselves by the $SU(2)$ case). 

They are 
based on the analogies with what is known for $4D$
theories.
Probably, the simplest way to calculate the $\beta$ function in the ordinary $4D$ SYM theory is
to use the supersymmetric background field technique \cite{Grisaru}. Fixing the gauge in supersymmetric way
brings about three adjoint chiral ghost multiplets. Two of them have nontrivial interactions with the quantum
fields, while the third one (the Nielsen-Kallosh ghost \cite{NK} ) is sterile in the quantum sense. But all three
ghosts interact with the background field in the same way as usual matter chiral multiplets, only the sign of their
contribution to the effective action is negative due to their ghost nature. And this is the only contribution 
as the loop of quantum vector superfield {\it vanishes} in this formalism. This gives the coefficient $-3$ 
in the ${\cal N} =1$ theory in SQED units. 
The ${\cal N} =2$ theory involves an extra ordinary adjoint chiral
multiplet, which gives $-3+1 = -2$. The ${\cal N} =4$ theory involves three such multiplets and $-3+3$ makes zero.

The  ${\cal N} =2$ and  ${\cal N} =4$ theories can also be described in the language of  ${\cal N} =2$
harmonic superfields. The ${\cal N} =4$ theory involves the gauge supermultiplet coupled to
 an adjoint matter hypermultiplet. The corresponding background field formalism was developed in \cite{Buch}. 
Also in this case
the contribution of the quantum gauge superfield loop vanishes. Also in this case, three ghost hypermultiplets:
two Faddeev-Popov ghosts  
 and the third Nielsen-Kallosh one, appear and contribute in the one-loop effective action. The difference with
the  ${\cal N} =1$ case is that the NK ghost is now {\it bosonic} rather than fermionic 
and it contributes
to $S^{\rm eff}$ with the same sign as the usual matter hypermultiplet. In ${\cal N} =2$ theory, the contribution is $2[(-2)_{\rm FP} + 1_{\rm NK} ] = -2$ in Abelian units.
The total  ${\cal N} =4$ contribution is proportional to 
$(-2)_{\rm FP} + 1_{\rm NK} + 1_{\rm MAT} \ = 0$.

Unfortunately, an  adequate background field supergraph technique in six dimensions has not been developed yet even for 
ordinary SYM theories, not speaking of higher-derivative ones.
 Still, one can suppose that this can be done. By analogy with 
 ${\cal N} =2$ $4D$ SYM, the full lagrangian
 should involve two fermionic Faddeev-Popov  and one bosonic Nielsen-Kallosh ghost  
hypermultiplets, and the net contribution to $S_{\rm eff}$ in $6D$  ${\cal N} =1$ theory should coincide with
the contribution of an adjoint hypermultiplet with the opposite sign.  This gives the result (\ref{g6Drenorm}).

Let us couple now this theory to an ordinary adjoint  matter hypermultiplet. The $\beta$ function should  vanish  in this case. 
 Such a theory should enjoy extended ${\cal N} =2$
supersymmetry (in the $6D$ sense) and should be liftable up to ten dimensions (note that the hypermutiplet
has 4 real scalar degrees of freedom for each color index and that $6+4 = 10$)  so that scalar fields enter the 
lagrangian in the same way as the components of the vector potential. 
One can conjecture that the
 extended supersymmetry  entails a nonrenormalization theorem, killing higher-loop contributions to 
the $\beta$ function. As a result, the theory is finite.

This theory involves a vacuum valley. Indeed,
 by $10D$ gauge invariance, the part of the lagrangian depending on the bosonic gauge field $A_M$ 
should have the form
 \be
\label{Lbos10D}
g^2 {\cal L}_{\rm bos} \ =\ \frac 12 \, {\rm Tr}\, \left\{ F_{MN} D^2 F_{MN} \right\}  + i\gamma {\rm Tr}
\left\{ F_{MN} F_{NP} F_{PM} \right \}\ , 
  \ee
 $M,N,P = 0,1,\ldots,9$ are 10--dimensional indices. The parameter $\gamma$ can be fixed by 
restricting  $M,N,P = 0,1,2,3$ and comparing (\ref{Lbos10D}) with the component expansion
of (\ref{N1SYMhder}). We find $\gamma = 2$. 
 Consider now the potential part of the lagrangian (\ref{Lbos10D})
in the scalar sector  $M,N,P = 6,7,8,9$,    
    \be
   \label{Vscal}  
  g^2V_{\rm scalar} \ = \ - \frac 12 \, {\rm Tr}\, \left\{  [A_P , [A_M, A_N]] [A_P,  [A_M,A_N]] \right \} 
 \nonumber \\   - 2 \, {\rm Tr}\, \left\{ [A_M, A_N] [A_N, A_P] [A_P, A_M] \right \}\ .
   \ee
We see that the static equations of motion $\delta V/\delta A_N = 0$ is satisfied, 
 provided $[A_M, A_N] = 0 $ for all $M,N = 6,7,8,9$.
Thus,  the vacuum   moduli space is characterized
by adjoint scalars living in the Cartan subalgebra of the Lie algebra, the situation familiar from the 
 ordinary SYM studies. Note that  the presence of the second term is important here.  
One would obtain without it some extra exotic valleys, like $A_{6,7,8} = 
c\tau^{1,2,3}, A_9 = 0$.

Now let vacuum be chosen (for $SU(2)$) as 
$A_M = c_M \tau^3$ and consider small fluctuations on this background.
Our main point is that nonzero expectation
values $c_M$ of vacuum scalars enter the lagrangian in the same way as the parameter $\Omega$
enters the lagrangian (\ref{L4}). One can illustrate this on a model example.  Consider 
the $2D$
lagrangian
  \be
 \label{L4gauge}
{\cal L} =  \frac 14 F_{\mu\nu} \Box F_{\mu\nu} + \bar \phi 
\left[(\partial_\mu - iA_\mu)^2 \right]^2 \phi 
  \ee
and reduce it to 0+1 dimensions.  By a gauge transformation, 
$A_0$ can be brought to zero.  After this, 
a nonzero expectation value of $A_1$ enters the reduced lagrangian exactly in
the same way as $\Omega$ in Eq.(\ref{L4}). 

For the actual lagrangian of interest (\ref{Lbos10D}), the situation
is roughly the same. Let us expand $A^a_N = c_N \delta^{3a} + \Phi^a_N$, where $c_N \delta^{3a}$ is the
classical vacuum field, gauge-rotated along the third color axis, and $\Phi^a_N$ is the quantum
fluctuation. To make the latter a genuine ``fast'' quantum fluctuation, which is charged
 with respect to the background 
and is not reduced to shifting the moduli
variable $c_N$ or gauge rotation, the constraints  $\Phi^3_N = 0, \, c_N \Phi^a_N = 0$ have to be imposed
(see e.g. \cite{Trieste}).

Consider only the zero momentum mode of $\Phi_N^a$ (we have seen that the zero mode
 is more viable to be unstable than
the others). The quadratic in   $\Phi_N^a$ part of the lagrangian has the form
$$\frac 1{2g^2} \Phi \left[ \frac {d^2 }{dt^2} + \vec{C}^2 \right]^2 \Phi \, ,$$
as in the toy model above.

Thus, linearized equations of motion are benign, they invoke no instability. It is not so easy to see, however,
whether the full nonlinear theory is stable or not. At the level $\sim \Phi^4$, there are 
different terms, some of
which involve a couple of extra time derivatives. Besides, there are the terms $\sim \Phi^6$. 
Different terms enter
with different signs and only a special study [like the one we performed for the toy model
(\ref{L4mix}) ] can determine whether the system exhibits the benign or malicious behavior. 
We are laying our  bets
on  the first, benign option.

Up to now, we only discussed the dynamics of fast charged fluctuations.
 The theory involves, however, also
neutral fields, which do not interact with the v.e.v.'s. There is no reason to believe that the
{\it zero modes} of these fields would not grow. In all probability, they would.
Let us consider the following model lagrangian
 \be
\label{L4y}
{\cal L} \ =\ \frac 12 (\ddot q  + y^2 q )^2  + \frac 12 \ddot y^2 - \frac 14 q^4 \ .
 \ee
We have just substituted the constant $\Omega$ in (\ref{L4}) by the dynamic variable $y(t)$ and
added the higher-derivative kinetic term for this variable. The potential part of the lagrangian
(\ref{L4y}) vanishes if $q=0, \ y = $ {\small anything}, which simulates the vacuum valley of the real lagrangian.  
The equations of motion are
  \be
\label{eqmotqy}
q^{(4)} + 2y^2 \ddot q + 2(\dot y^2 + y \ddot y) q + 4 y \dot y \dot q + y^4 q - q^3 &=& 0\, , \nonumber \\
y^{(4)} + 2 y q \ddot q + 2y^3 q^2 &=& 0
  \ee

\begin{figure}[h]
   \begin{center}
 \includegraphics[width=4.0in]{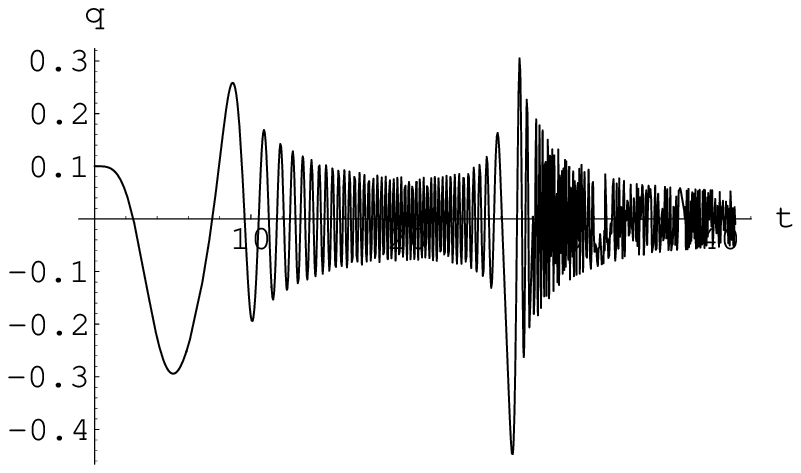}
 \includegraphics[width=4.0in]{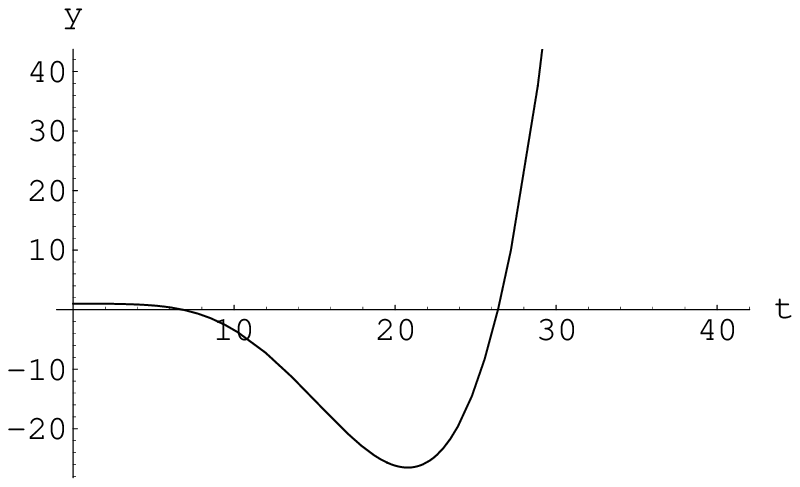}
        \vspace{-6mm}
    \end{center}
\caption{\small $q(t)$ and $y(t)$ for the system (\ref{eqmotqy}) with the initial conditions (\ref{incond}).}
\label{traqy}
\end{figure}

 The numerical solutions to these equations with initial conditions
   \be
  \label{incond}
 y(0) = 1, \ \ \dot y(0) = \ddot y(0) = y^{(3)} = 0;\ \ \ 
q(0) = 0.1, \ \ \dot q(0) = \ddot q(0) = q^{(3)} = 0
   \ee
are shown in Fig.\ref{traqy}.
 We see that the ``charged'' variable $q$ does not grow while 
instability with respect to drifting along the valley is present, indeed. It shows
up however small the initial value $q(0)$ is.
Only the zero mode of the valley scalar field grows and this again could be associated with inflation.
But as we are in six dimensions now, we do not dare to speculate further in this direction.
Note that this instability does not spoil the unitarity of the scattering matrix. Asymptotic
states have always nonzero momenta and the absence of instabilities with respect to small
perturbations in this sector means that the dispersion law for the asymptotic states has a normal
form with only real solutions for the energies. There are no malicious ghosts !
     \footnote{A tacit assumption here was that the physical asymptotic states coincide with the
    states extracted from the fundamental lagrangian. { This} {\it is} the place where the finiteness
     of our theory is used. We know that  an asymptotically free theory, like (\ref{SYMhder}), involves
dimensional transmutation and confinement phenomenon. In this case, it is not quite clear for us whether
the benign nature of ghosts can be preserved.}
Bearing in mind this and renormalizability of the theory,
one can conclude that
 perturbation theory is absolutely OK there. The  problems associated with the growth of the zero
``inflationary'' mode and with
the danger of collapse due to large nonperturbative perturbations remain. 
But, as was 
mentioned before, the latter is an intrinsic problem for {\it any}  theory that includes gravity, and we
do not pretend to know how to solve it.

We started this article with discussing  higher-derivative gravity in four dimensions, 
in particular, conformal supergravity.
 It is natural to ask now
whether our mechanism for taming the ghosts works there, as it works in higher-derivative $6D$ SYM.
 The finite ${\cal N} =4$ conformal supergravity involves the gravity
supermultiplet coupled to  the ${\cal N} =4$ SYM multiplet. There are  vacuum valleys associated with the scalars
from SYM supermultiplet and also the valleys associated with certain scalar superpartners $E_{(ij)}$ 
($i,j = 1,\ldots, 4$ is the SUSY flavour index) to the graviton. Nonzero vacuum expectation values of these
scalars provide for the induced Einstein term $\sim R$ in the action.\,\footnote{ Here the Einstein term appears at
the classical level. Long time ago Sakharov noticed that the Einstein term in the effective action
is induced in large class of theories if taking loops into account \cite{Sakharov}. In particular, it is
generated in higher-derivative gravity and conformal supergravity   \cite{ja+Adler}.}
 On the other hand, the cosmological term is not generated. Indeed, a nonzero induced cosmological term would mean
a nontrivial {\it potential} along the valley. The whole point, however, is that such potential is absent.

If we choose the positive sign of the Weyl term $ C_{\mu\nu\rho\sigma}^2$
in the Minkowski action, the scalars $E$ enter with the usual sign $\sim (\partial_\mu E)^2$. The fields
$E$ couple to the metric in a conformally invariant way
$$\sim \frac 12 (\partial_\mu E)^2 - \frac 1{12} R E^2\ . $$
We see that a  nonzero vacuum expectation value $\langle E \rangle $ generates the Einstein term with the {\it negative}
coefficient \cite{Fradkin}. The expectation values of the  scalar superpartners of SYM gauge fields also give a 
negative contribution to the effective Einstein constant. 
In other words, the effective lagrangian in the gravity sector is 
$  {\cal L}_{\rm grav}  \sim C_{\mu\nu\rho\lambda}^2 - \kappa^2 R$. The corresponding dispersive equation
$p^2(p^2 - \kappa^2/2 ) = 0$
has only real solutions and the linearized theory {\it is} benign ! One notices that the negative coefficient
of $R$ is crucial here. The theory $C^2 + \kappa^2 R$ would be badly malicious.

Of course, a negative gravity constant
is not what one observes in experiment. Apples fall down to Earth rather than
fly up to sky. In principle, this can be handled by 
reversing the overall sign of the lagrangian, which is a pure convention. The price one has to pay for that is 
rather high, however: all usual spin 0 and spin 1 fields become ghosts 
(see \cite{Fradkin} for more detailed discussion). Thus, a direct phenomenological relevance
of the finite ${\cal N} = 4$ conformal supergravity is not yet clear.       

Leaving aside phenomenology, a legitimate question to ask is whether the ${\cal N} =4$ conformal supergravity 
is benign or malicious, whether the vacuum is stable or not there with respect to 
small spatially-dependent fluctuations of the metric. To answer this, one should study again the 
full nonlinear classical 
field equations in the vicinity of the vacuum. It is not a simple but feasible   study.

\section{Conclusions.}
Our main observations are the following.
\begin{enumerate}
\item All the trouble associated with ghosts is clearly seen at the classical level. It {\it suffices} to
study classical equations of motion to understand whether the trouble is serious or it can be cured, and to 
what extent. 
\item We have constructed toy mechanical higher-derivative models where this trouble is present in a weakened
benign form: vacuum is stable with respect to small perturbations, thereby perturbation theory is well defined,
but, being shaken by a large perturbation, the system might collapse.
 \item We argued that certain finite higher-derivative field theories, involving nontrivial vacuum moduli
spaces, namely, the $6D$  ${\cal N} = 2$ higher-derivative SYM theory and maybe also a finite version of
$4D$  ${\cal N} =4$ conformal supergravity, have similar properties. They are stable with respect to 
small perturbations with nonzero momenta. If so, this makes perturbation theory in these theories perfectly
well defined. In particular, $S$-matrix is unitary  order by order.
 \item These theories always involve, however, 
instability of certain zero-momentum field modes associated with drifting
along the vacuum valley. Large coherent field thereby produced is a prerequisit for inflation 
and one can speculate that the physical inflation is, indeed, 
associated with such an instability in a higher-derivative theory. 

\end{enumerate}

Is the TOE  a form of higher-derivative field theory, indeed ? We do not know, but maybe it is. We would sound 
more enthusiastically if  the theories discussed in this paper could be consistently formulated in
Euclidean space. Unfortunately, for 
{\it supersymmetric}    higher-derivative theories, this is not the case. Their Euclidean action involves
the terms of different sign. This is true for $6D$ SYM [see Eq.(\ref{phhdercomp})], this is true for conformal
supergravity, and this seems to be a family feature of all such theories. In other words, a sensible 
nonperturbative definition of path integral in these theories is absent.   

On the other hand, purely bosonic  
 higher-derivative theories with positive definite Euclidean action exist. Such is the pure Weyl gravity.  
The Euclidean
action of the toy model (\ref{L4dot}) is also positive definite (as is written, it is rather  negative definite, 
but this can be fixed by reversing the overall
sign of the lagrangian). The model (\ref{L4dot}) suffers from collapse, 
but all Euclidean correlators are well defined there and the consistent {\it nonperturbative} 
quantum theory, based on the lagrangian
(\ref{L4dot}), can probably  be constructed. We are not ready, however, 
to discard supersymmetry as a guiding
principle for TOE. In particular, the mechanism for eradicating the malicious 
nature of ghosts studied
in this paper depends crucially on supersymmetry. Non-supersymmetric theories are not finite, 
vacuum valleys are distorted  there by quantum correction, etc

The last remark is that the conformal $6D$  ${\cal N} = 2$ theory considered in Sect.3 has an additional nice feature.
Assuming, as we did when writing (\ref{Lbos10D}), that the bosonic part of the action can be obtained from the action of 
the  $6D$  ${\cal N} = 1$ theory by lifting it to 10 dimensions (to derive accurately 
(\ref{Lbos10D}) in the framework of the $6D$
superfield formalism it is a separate not  yet solved problem), one can be convinced \cite{ISZ} 
that familiar BPS monopole configurations (they represent 2-branes being localized in three dimensions and not depending
on two remaining spatial coordinates) satisfy the equations of motion. A new remarkable feature compared to what is known 
for $4D$ extended SYM theories is that the energy density of such a brane vanishes identically. If thinking about this brane
as of a (2+1)-dimensional Universe, zero energy density implies the zero value for
 the induced cosmological term in this Universe ! For sure, we live in the (3+1)-dimensional Universe, not (2+1) dimensional
one and the model we are discussing is not phenomenologically acceptable, but one can still 
speculate that the absence (or a very
small value) of the cosmological term in our Universe might have a similar explanation.
\footnote{Another wild speculation is the following. Suppose that for some reason only the first term of the potential
(\ref{Vscal}) is present and the second is not. Then, as was mentioned, a new {\it non-Abelian} scalar vacuum valley like
$A_{6,7,8} = c\tau^{1,2,3}$ is present. In such a case, $\pi_1$ of the corresponding coset is nontrivial and 
Abrikosov string solutions might exist \cite{MMY}. These strings become $4D$ branes in the $6D$ bulk \cite{Pol}...}  

One thing is { quite} clear: more work in this direction is necessary.

I am indebted to E. Ivanov and A. Tseytlin for  illuminating discussions and many valuable remarks 
before and after reading the manuscript and to I. Buchbinder  for very useful comments.

\end{document}